\documentclass[12pt]{article}
\usepackage{latexsym}
\usepackage{amsfonts}
\usepackage{epsfig}

\catcode`\@=11

\def\section{\@startsection{section}{1}{\z@}{3.5ex plus 1ex minus
 .2ex}{2.3ex plus .2ex}{\bf}}

\def\thesubsection{\arabic{section}.\arabic{subsection}}
\renewcommand{\subsection}[1]{\addtocounter{subsection}{1}
\vspace{2.5mm}\par\noindent {\it \thesubsection . #1}\par
 \vspace{0.5mm} }
\catcode`\@=12
\mathchardef\varGamma="0100 \mathchardef\varDelta="0101
\mathchardef\varTheta="0102 \mathchardef\varLambda="0103
\mathchardef\varXi="0104 \mathchardef\varPi="0105
\mathchardef\varSigma="0106 \mathchardef\varUpsilon="0107
\mathchardef\varPhi="0108 \mathchardef\varPsi="0109
\mathchardef\varOmega="010A
\def\bfone{\relax{\rm 1\kern-.35em 1}}
\DeclareFontFamily{U}{rsf}{} \DeclareFontShape{U}{rsf}{m}{n}{
  <5> <6> rsfs5 <7> <8> <9> rsfs7 <10-> rsfs10}{}
\DeclareMathAlphabet\Scr{U}{rsf}{m}{n}
\begin{document}
\begin{titlepage}
\thispagestyle{empty}
\begin{flushright}
\hfill{CERN-PH-TH/2004-212} \\\hfill{DESY-04-211}
\end{flushright}
\vspace{35pt}
\begin{center}{ \LARGE{\bf
Gauging the Heisenberg algebra of special quaternionic
manifolds}}\\
 \vspace{60pt} {\bf R. D'Auria$^\S$, S.
Ferrara$^\sharp$, M. Trigiante$^\S$, S. Vaul\`a$^\natural$}\\
\vspace{15pt}
$\S${\it Dipartimento di Fisica, Politecnico di Torino \\
C.so Duca degli Abruzzi, 24, I-10129 Torino\\
Istituto Nazionale di Fisica Nucleare, Sezione di Torino,
Italy}\\[1mm] {E-mail: riccardo.dauria@polito.it,
mario.trigiante@polito.it}\\
\vspace{15pt} $\sharp${\it CERN, Physics Department, CH 1211
Geneva 23, Switzerland.}\\ {\it INFN, Laboratori Nucleari di
Frascati, Italy}\\[1mm] {E-mail:
sergio.ferrara@cern.ch}\\
 \vspace{15pt}
$\natural${\it DESY, Theory Group\\
Notkestrasse 85, Lab. 2a D-22603 Hamburg, Germany\\
II. Institut f\"ur Theoretische Physik,\\
Luruper Chaussee 149, D-22761 Hamburg, Germany}\\[1mm] {E-mail:
silvia.vaula@desy.de}
\vspace{15pt}
\begin{abstract}
We show that in $N=2$ supergravity, with a special quaternionic
manifold of (quaternionic) dimension $h_1+1$ and in the presence
of $h_2$ vector multiplets, a $h_2+1$ dimensional abelian algebra,
intersecting the $2h_1+3$ dimensional Heisenberg algebra of
quaternionic isometries, can be gauged provided the $h_2+1$
symplectic charge--vectors $V_I$, have vanishing symplectic
invariant scalar product $V_I\times V_J=0$.
 For compactifications on Calabi--Yau three--folds with Hodge
 numbers $(h_1,h_2)$ such condition generalizes the half--flatness
 condition as used in the recent literature.
 We also discuss non--abelian extensions of the above
 gaugings and their consistency conditions.
\end{abstract}
\end{center}
\end{titlepage}
\newpage
\baselineskip 6 mm
\section{The Heisenberg algebra}
It is well known \cite{cfg,fs} that the moduli space of a
Calabi--Yau compactification of Type II string theory is a product
of a special quaternionic manifold ${\Scr M}_{SQ}$ of quaternionic
dimension $h_1+1$ and a special K\"ahler manifold ${\Scr M}_{SK}$
of complex dimension $h_2$ where $h_1=h_{(2,1)},\, h_2=h_{(1,1)}$
for Type IIA and the reverse for
Type IIB.\\
The special quaternionic geometry has some general properties
\cite{fs,dwvp1,dwvp2}, i.e. the $2h_1+3$ coordinates which
describe $2h_1+2$ R--R scalar fields and the $a$ scalar field dual
to the antisymmetric tensor field $B_{\mu\nu}$ \cite{dsv},
parametrize, in the  ``solvable description'' of the manifold
\cite{adfft}, a Heisenberg algebra of the form:
\begin{eqnarray}\label{Heis}
\left[X^\Lambda,\, Y_\Sigma\right]&=&\delta^\Lambda{}_\Sigma\, {\Scr Z}\,\,\,;\,\,\,\,\,
\Lambda=0,\dots, h_1\,,\nonumber\\
\left[X^\Lambda,\, X^\Sigma\right]&=&\left[Y_\Lambda,\,
Y_\Sigma\right]=\left[X^\Lambda,\,{\Scr
Z}\right]=\left[Y_\Lambda,\, {\Scr Z}\right]=0\,.
\end{eqnarray}
In Calabi--Yau compactifications the generators
$X^\Lambda,\,Y_\Sigma$ in (\ref{Heis}) are parametrized by the RR
real scalars which in Type IIA come from the internal components
of the complex 3--form $A_{(3)}$ \cite{bcf,bghl,lm}:
\begin{eqnarray}
\{\tilde{\zeta}_\Lambda,\,\zeta^\Lambda\}&\rightarrow
\{A_{ijk},\,A_{i\bar{j}\bar{k}},\,A_{\bar{i}\bar{j}\bar{k}},\,A_{\bar{i}jk},\,\}\,.
\end{eqnarray}
while Type IIB they originate from the 2--form and 4--form
cohomology:
\begin{eqnarray}
\{\tilde{\zeta}_\Lambda,\,\zeta^\Lambda\}&\rightarrow
\{C,\,C_{\bar{i}j \,\bar {l}k},\,C_0,\,C_{i\bar{j}}\}\,,
\end{eqnarray}
where $C$ is the dual of $C_{\mu\nu}$.\par The universal
hypermultiplet contains, besides the dilaton and the $a$ field
which parametrizes the generator ${\Scr Z}$ in (\ref{Heis}),he
$\Lambda=0$ component of the above coordinates, namely $\{{\rm
Re}A_{ijk},\,{\rm Im}A_{ijk}\}$ in Type IIA and $\{C_0,\, C\}$ in
Type IIB. In each case such multiplets parametrize ${\Scr
M}_U={\rm SU}(1,2)/{\rm U}(2)\subset {\Scr M}_{SQ}$. Under the
group of motions generated by the Heisenberg algebra the scalar
fields $\tilde{\zeta}_\Lambda,\,\zeta^\Lambda$ transform as
follows \cite{fs}:
\begin{eqnarray}
\delta\zeta^\Lambda&=&u^\Lambda\,\nonumber\\
\delta\tilde{\zeta}_\Lambda&=& v_\Lambda\,\nonumber\\
\delta a &=&
w+u^\Lambda\tilde{\zeta}_\Lambda-v_\Lambda\zeta^\Lambda\,.
\label{transf}\end{eqnarray} Noting that
$\delta(\tilde{\zeta}_\Lambda\zeta^\Lambda)=u^\Lambda\tilde{\zeta}_\Lambda+v_\Lambda\zeta^\Lambda$
we may redefine $a$ in such a way that one of the two
scalar--dependent terms in $\delta a$ is eliminated.
\section{The gaugings}
Let us define a gauge algebra through the following infinitesimal
field transformations:
\begin{eqnarray}
\delta A_\mu^I&=&\partial_\mu \lambda^I\,,\nonumber\\
\delta \zeta^\Lambda&=&a_I{}^\Lambda\, \lambda^I\,,\nonumber\\
\delta \tilde{\zeta}_\Lambda&=&b_{I\Lambda}\, \lambda^I\,,\nonumber\\
\delta a &=&c_I\,
\lambda^I+(a_I{}^\Lambda\,\tilde{\zeta}_\Lambda-b_{I\Lambda}\,\zeta^\Lambda)\,\lambda^I\,,
\end{eqnarray}
where $I=0,\dots , h_2$, $h_2$ being the number of vector
multiplets. Note that no relation exists between $h_1,\,h_2$ so
that the above algebra is not in general contained in the
Heisenberg algebra.\par The covariant derivatives read:
\begin{eqnarray}
D_\mu \zeta^\Lambda&=&\partial_\mu\zeta^\Lambda-a_I{}^\Lambda\,
A^I_\mu\,,\nonumber\\
D_\mu
\tilde{\zeta}_\Lambda&=&\partial_\mu\tilde{\zeta}_\Lambda-b_{I\Lambda}\,
A^I_\mu\,,\nonumber\\
D_\mu a&=&\partial_\mu
a-(a_I{}^\Lambda\,\tilde{\zeta}_\Lambda-b_{I\Lambda}\,\zeta^\Lambda)\,
A^I_\mu-c_I\,A^I_\mu\,.
\end{eqnarray}
One can verify that:
\begin{eqnarray}
\delta (D_\mu \zeta^\Lambda)&=&\delta(D_\mu
\tilde{\zeta}_\Lambda)=0\,,\nonumber\\ \delta(D_\mu
a)&=&(a_I{}^\Lambda\,D_\mu\tilde{\zeta}_\Lambda-b_{I\Lambda}\,D_\mu\zeta^\Lambda)\,\lambda^I\,,
\end{eqnarray}
where in order to derive the last equation we required requires
the following condition:
\begin{eqnarray}
c_{IJ}&\equiv &b_{I\Lambda}\, a_J{}^\Lambda-b_{J\Lambda}\,
a_I{}^\Lambda\,=0\,,\label{cocy}
\end{eqnarray}
which we shall characterize in the sequel as a ``cocycle''
condition of the Lie algebra. If we consider $\{
a_J{}^\Lambda,\,b_{I\Lambda}\}$ to be the $2\,h_1+2$ components of
a symplectic vector $V_I$, condition (\ref{cocy}) can be rephrased
as the vanishing of the symplectic scalar product $V_I\times
V_J=0$. Such condition is also equivalent to the closure of the
abelian gauge algebra whose generators $\{T_I\}$ are:
\begin{eqnarray}\label{embe1}
T_I&=&b_{I\Lambda}\, X^\Lambda+a_I{}^\Lambda\, Y_\Lambda+c_I\,
{\Scr Z}\,\,\,;\,\,\,\,\, \left[T_I,\,T_J\right]=0\,.
\end{eqnarray}
\section{Gauging of special quaternionic $\sigma$--model}
There is an elegant way of writing the RR scalars in the
quaternionic manifold in terms of the symplectic section:
\begin{eqnarray}
Z&=&\left(\matrix{\zeta^\Lambda\cr
\tilde{\zeta}_\Lambda}\right)\,,
\end{eqnarray}
and the symplectic (symmetric) matrix ${\Scr M}$ \cite{cdf}:
\begin{eqnarray}
{\Scr M}&=&\left(\matrix{\bfone & -{\rm Re}({\Scr N})\cr 0 &
\bfone }\right)\left(\matrix{{\rm Im}({\Scr N}) & 0\cr 0 & {\rm
Im}({\Scr N})^{-1} }\right)\left(\matrix{\bfone &0\cr -{\rm
Re}({\Scr N})& \bfone}\right)\,.
\end{eqnarray}
Indeed the kinetic term \cite{fs} is given by:
\begin{eqnarray}
&&K_{a\bar{b}}\, \partial_\mu z^a\partial^\mu
\bar{z}^{\bar{b}}-\frac{1}{4\,\phi^2}\,(\partial\phi)^2-
\frac{1}{4\,\phi^2}\,(\partial a -Z\times
\partial Z)^2-\frac{1}{2\,\phi}\,\partial_\mu Z\, {\Scr
M}\,\partial^\mu Z\,.
\end{eqnarray}
Note that invariance under the Heisenberg algebra with symplectic
parameters:
\begin{eqnarray}
\Theta&=&\left(\matrix{u^\Lambda\cr v_\Sigma}\right)\,,
\end{eqnarray}
is manifest sice
\begin{eqnarray}
\delta a&=&w+\Theta\times Z\,\,;\,\,\,\delta
Z=\Theta\,\,\Rightarrow \,\,da-Z\times dZ \,\,\mbox{invariant}\,.
\end{eqnarray}
The gauging of the non linear $\sigma$--model goes as follows. We
consider an abelian $h_2+1$ dimensional gauge group whose
embedding in the Heisenberg algebra is described by $h_2+1$
symplectic charge-vectors
\begin{eqnarray}
V_I&=&\left(\matrix{a_I{}^\Lambda\cr b_{I\,\Lambda}}\right)\,,
\end{eqnarray}
and whose connection $U$ is expressed in terms of the $h_2+1$
vector fields $A^I_\mu$ as follows:
\begin{eqnarray}
U&=&A^I\,V_I\,,\\
\delta U&=& d\Theta\,,
\end{eqnarray}
$\Theta$ being now $\Theta=\lambda^I\,V_I$, where $\lambda^I$ are
the gauge parameters. The covariant derivative of $Z$ is then
\begin{eqnarray}
DZ&=&dZ-U\,,
\end{eqnarray}
and the covariant derivative of $a$ reads
\begin{eqnarray}
Da&=&da-U\times Z\,,
\end{eqnarray}
since $\delta a =\Theta\times Z$.\par If we transform $Da$ we
obtain
\begin{eqnarray}
\delta(Da)&=&\Theta\times dZ-U\times \Theta=\Theta\times (dZ+U)\,,
\end{eqnarray}
which is not $\Theta\times DZ$ since $DZ=dZ-U$. Therefore closure
implies $\Theta\times U=0$ which is equivalent to condition
(\ref{cocy}) since:
\begin{eqnarray}
\Theta\times U&=&2\,a_{[I}{}^\Lambda\,
b_{J]\,\Lambda}\,\lambda^I\, A^J\,.
\end{eqnarray}
If $\Theta\times U=0$ we can write the RR sector of the gauged
Lagrangian
\begin{eqnarray}
&&- \frac{1}{4\,\phi^2}\,(D a -Z\times D
Z)^2-\frac{1}{2\,\phi}\,D_\mu Z\, {\Scr M}\,D^\mu Z\,.
\end{eqnarray}
Upon addition of the minimal coupling $\partial a -c_I\, A^I$ to
the covariant derivative of $a$ the vector boson mass matrix
$M^2_{IJ}$ will read:
\begin{eqnarray}
M^2_{IJ}&=&\frac{1}{2\,\phi^2}\,(c_I-2\,Z\times
V_I)\,(c_J-2\,Z\times V_J) +\frac{1}{\phi}\,V_I\,{\Scr M}V_J\,.
\end{eqnarray}
\section{Non--abelian gauging}
Let us now see what are the requirements which have to be
satisfied in order to embed a non--abelian gauge algebra in the
Heisenberg algebra.\\
 Quite generally we introduce a non--abelian
gauge algebra defined by:\begin{equation}\label{gaugealg}
    [T_I,\,T_J]=f^K_{\phantom {T}IJ}T_K\,.
\end{equation}
Using the embedded expression for the gauge algebra generators
given in equation (\ref{embe1}), the embedding condition
\begin{equation}\label{embe2} f^K_{\phantom {T}IJ}T_K=c_{IJ}{\Scr
Z}\,,
\end{equation}
implies the following relations:
\begin{eqnarray}\label{cond1}
f^K_{\phantom {T}IJ}c_K&=&c_{IJ}\,, \\
\label{cond2}f^K_{\phantom {T}IJ}b_{K\Lambda}&=&0\,,\\
\label{cond3}f^K_{\phantom {T}IJ}a_K{}^{\Lambda}&=&0\,.
\end{eqnarray}
In terms of the Lie algebra cohomology equation (\ref{cond1})
means that $c_{IJ}$ is a non trivial cocycle of the gauge algebra,
while (\ref{cond2}) and (\ref{cond3}) imply that $b_{I\Lambda}$
and $a_I{}^{\Lambda}$ are coboundaries. When $c_{IJ}=0$ the
cohomology is trivial and we are in the case of the abelian gauge
algebra discussed in the previous section. Since the algebra
(\ref{embe2}) contains a central charge it is non-semisimple and
according to a theorem of Lie algebra cohomology we may have a non
trivial cocycle $c_I$ in the adjoint representation of the algebra
(this would be impossible if the gauge algebra were semisimple
since in that case the only non trivial cocycle should be in the
trivial representation of the algebra). In fact a solution of
conditions (\ref{cond1}),(\ref{cond2}),(\ref{cond3}) may be found
as follows. We first consider the case in which $h_2+1=2 h_1+3$,
so that the number of vector matches the dimension of the
Heisenberg algebra. The gauge generators $T_I$ decompose in the
following way:
\begin{eqnarray}
\{T_I\}&=&\{T_\Lambda,\,T^\Lambda,\,T_0\}\,.
\end{eqnarray}
The charge matrices are chosen to be
\begin{eqnarray}
b_{0\Lambda}&=&b_{\Sigma\Lambda}=0\,\,;\,\,\,b^\Sigma{}_\Lambda=b_\Lambda\,\delta^\Sigma{}_\Lambda\,,\nonumber\\
a_{0}{}^{\Lambda}&=&a^{\Sigma\Lambda}=0\,\,;\,\,\,a_\Sigma{}^\Lambda=a^\Lambda\,\delta_\Sigma{}^\Lambda\,,\\
\end{eqnarray}
The cocycle condition (\ref{cond1}) becomes
\begin{eqnarray}
c_0\,f_{\Lambda}^0{}^\Sigma&=&(b_\Lambda
a^\Lambda)\,\delta_\Lambda{}^\Sigma\,,
\end{eqnarray}
with no summation over the index $\Lambda$. Conditions
(\ref{cond2}), (\ref{cond3}) are manifestly verified. If
$h_2>2\,(h_1+1)$ we may apply the above construction to $2\,
h_1+3$ vectors while the remaining $h_2-2\,(h_1+1)$ vectors stay
spectators. Viceversa, if $h_2<2\,(h_1+1)$ we can select a
Heisenberg subalgebra with $\bar{h}_1=\frac{h_2}{2}-1$ and apply
to it the construction described above.
\section{Gauging and half--flatness}
Let us consider Calabi--Yau compactifications on a
\emph{half--flat} manifold \cite{halfflat}. For the kind of
manifolds considered in \cite{gm,glmw}, in the absence of fluxes,
we have the following couplings in Type IIA and IIB theories
\begin{eqnarray}
\mbox{IIA}&&a_I{}^\Lambda=0\,\,\,;\,\,\,\,b_{I\Lambda=0}=\epsilon_I\,;\,\,\,\,(\mbox{0
otherwise})\,,\\
\mbox{IIB}&&a_I{}^\Lambda=0\,\,\,;\,\,\,\,b_{I=0\Lambda}=\epsilon_\Lambda\,;\,\,\,\,(\mbox{0
otherwise})\,,
\end{eqnarray}
and the cocycle condition (\ref{cocy})is identically satisfied
(recall that, according to our notations, in Type IIA $I=0,\dots,
h_{1,1}$ and $\Lambda=0,\dots, h_{2,1}$ while in Type IIB
$I=0,\dots, h_{2,1}$, $\Lambda=0,\dots, h_{1,1}$). Note that we
use the same symbols to denote the charges
$a_I{}^\Lambda,\,b_{I\Lambda}$ in Type IIA and IIB theories
although they are described in the two cases by different matrices
with different dimensions. If we turn on a NS 3--form flux in Type
IIA theory we have $a_{I=0}{}^\Lambda=p^\Lambda\neq 0$ and
$b_{I=0\Lambda}=q_\Lambda\neq 0$, and then, on a half--flat
manifold we should also have a non vanishing $a_{I}{}^\Lambda$
since the cocycle condition requires:
\begin{eqnarray}
a_{I}{}^\Lambda \, b_{J\Lambda}&=&a_{J}{}^\Lambda \,
b_{I\Lambda}\Rightarrow q_0\,a_I{}^0=p^0\, \epsilon_I\,.
\end{eqnarray}
On the Type IIB side, if we turn on an electric NS 3--form flux
 we get a covariant derivative of the type \cite{m}:
\begin{eqnarray}
D_\mu\tilde{\zeta}_0=\partial_\mu\tilde{\zeta}_0 -q_{I}\,
A^I_\mu\,,
\end{eqnarray}
where $q_I$ is the \emph{electric} flux $b_{I\,\Lambda=0}$. If
Type IIB background is half--flat \cite{gm} we also have
$b_{I=0\,\Lambda}=\epsilon_{\Lambda}\neq 0$. In this case, as
expected, $b_{0\,0}=\epsilon_0=q_0$. For the \emph{magnetic} NS
3--form flux the correspondence is non--local.\par The abelian
gauging of the Heisenberg algebra discussed in the previous
sections therefore generalizes the results on
flux--compactifications on half--flat manifolds as discussed in
the literature \cite{gm,glmw}, to arbitrary values of
$I,\,\Lambda$. Consistency always requires in the ``dual
theories'' the cocycle condition to be satisfied:
\begin{eqnarray}
a_{[I}{}^\Lambda\,
b_{J]\Lambda}&=&0\,\,\,\,;\,\,\,\,\,(I,J=0,\dots,h_2\,;\,\,\Lambda=0,\dots,
h_1)\,.\nonumber
\end{eqnarray}
Mirror symmetry on the other hand implies
\begin{eqnarray} b^{(B)}{}_{I\,\Lambda}&=&(b^{(A) T})_{
I\Lambda}\,\,\,;\,\,\,\,(I=0,\dots,h_{2,1}\,;\,\,\Lambda=0,\dots,
h_{1,1})\,,\nonumber\end{eqnarray}
 In Type IIA theory we can interpret the parameters $a_I{}^\Lambda,\,b_{I\,\Lambda}$
  of the gauging in terms of the following
  deformation of the Calabi--Yau cohomology \cite{glmw,gm}:
\begin{eqnarray}
d\alpha_\Lambda&=&b_{i\Lambda}\,
\omega^i\,\,;\,\,\,\,d\beta^\Lambda=a_i{}^\Lambda\,
\omega^i\,,\nonumber\\
d\omega_i&=&a_i{}^\Lambda\,\alpha_\Lambda-b_{i\Lambda}\,\beta^\Lambda\,,\nonumber\\
\omega_i&\in& H^{(1,1)}\,\,;\,\,\,\,\omega^i\in
H^{(2,2)}\,\,;\,\,\,\,i=1\dots, h_{1,1}\,.\label{eqs1}
\end{eqnarray}
in the presence of a non trivial NS flux:
\begin{eqnarray}
\hat{H}_{(3)}&=&dB_{(2)}+d(b^i\,\omega_i)-a_{0}{}^{\Lambda}\,
\alpha_\Lambda+b_{0\Lambda}\beta^\Lambda\,.
\end{eqnarray}
 Integrability on the cohomology side gives:
\begin{eqnarray}
d\omega^i&=&0\,\,\,;\,\,\,\,\,d^2\omega_i=-(a_i{}^\Lambda\,b_{j\Lambda}-a_{j}{}^\Lambda\,b_{i\Lambda})\,
\omega^j=0\,,\label{cocy1}
\end{eqnarray}
while on the NS flux it implies
\begin{eqnarray}
d\hat{H}_{(3)}&=&0\,\,\,\Rightarrow\,\,\,\,\,(a_0{}^\Lambda\,b_{j\Lambda}-a_{j}{}^\Lambda\,b_{0\Lambda})\,
\omega^j=0\,.\label{cocy2}
\end{eqnarray}
Conditions (\ref{cocy1}),(\ref{cocy2})are equivalent to the
cocycle condition (\ref{cocy}).\par One can show that the
definition of the $a_I{}^\Lambda,\,b_{I\,\Lambda}$ given in
(\ref{eqs1}) is consistent. Indeed, for instance, on one hand we
can write:
\begin{eqnarray}
\int d\alpha_\Lambda\wedge \omega_i&=&-\int \alpha_\Lambda\wedge
(a^\Sigma{}_{j}\,\alpha_\Sigma-b_{j\Sigma}\,\beta^\Sigma)=b_{j\Lambda}\,,
\end{eqnarray}
while on the other hand we have:
\begin{eqnarray}
b_{i\Lambda}\,\int \omega^i\wedge \omega_j&=&b_{j\Lambda}\,.
\end{eqnarray}
By performing a compactification on such \emph{half--flat}
Calabi--Yau in the presence of a NS flux we have
\begin{eqnarray}
d\hat{A}&=& dA_0\,,\nonumber\\
d\hat{B}_{(2)}&=& dB_{(2)}+db^i\wedge \omega_i-b^i
\,(a_i{}^\Lambda\,\alpha_\Lambda-b_{i\Lambda}\,\beta^\Lambda)\,,
\nonumber\\
d\hat{C}_{(3)}&=& d\tilde{A}^i\wedge \omega_i-(\zeta^\Lambda\,
b_{i\Lambda}-\tilde{\zeta}_\Lambda\,
a_i{}^\Lambda)\,\omega^i-(d\zeta^\Lambda-a_i{}^\Lambda\,\tilde{A}^i)\wedge
\alpha_\Lambda+\nonumber\\&&(d\tilde{\zeta}_\Lambda-b_{i\Lambda}\,\tilde{A}^i)\wedge
\beta^\Lambda\,,\nonumber\\
\hat{F}_{(4)}&=&d\hat{C}_{(3)}+\hat{H}_{(3)}\wedge
A^0=dA^i\wedge\omega_i-b^i\,dA^0\wedge
\omega_i-(d\zeta^\Lambda-a_I{}^\Lambda\,
A^I)\wedge\alpha_\Lambda+\nonumber\\&&(d\tilde{\zeta}_\Lambda-b_{I\,\Lambda}\,
A^I)\wedge\beta_\Lambda-(\zeta^\Lambda\,
b_{i\Lambda}-\tilde{\zeta}_\Lambda\,
a_i{}^\Lambda)\,\omega^i+dB_{(2)}\wedge A^0\,,\\
A^i&=&\tilde{A}^i+b^i\,A^0\,.\nonumber
\end{eqnarray}
So we obtain the correct gauging of the Ramond isometries. The
covariant derivative of the scalar field $a$ dual to $B_{\mu\nu}$
is obtained from the topological term in the IIA ten--dimensional
action as in \cite{glmw}.
\section{Conclusions}
In this note we have studied the gauging of the Heisenberg algebra
which is common to all special quaternionic manifolds, and proved
that, for an abelian gauge algebra, it requires a vanishing
cocycle condition (i.e. that a certain Lie algebra cocycle be
trivial). This gauge algebra, as it appears in Calabi--Yau
compactification with fluxes or/and half--flat manifolds,
corresponds to the gauging of isometries acting on RR scalars and
the (dual of) the NS 2--form. The symplectic structure exhibited
by the RR scalars embedded in a special quaternionic manifold
suggests the general form of the gauging and a mirror relation
when switching to the Heisenberg algebra of the mirror theory. It
is suggestive that, if this is done, new couplings are predicted
that do not usually appear in the perturbative formulation of Type
IIA and Type IIB theories. The general gauging of the Heisenberg
algebra also induces a scalar potential which, in some particular
cases, has been studied in \cite{gm} and \cite{glmw}, and whose
general properties are under investigation.

\section{Acknowledgements}
 R.D. and M.T. would like to thank the Physics
Department of CERN, where part of this work was done, for its kind
hospitality.
\par
 Work
supported in part by the European Community's Human Potential
Program under contract HPRN-CT-2000-00131 Quantum Space-Time, in
which R. D'A. is associated to Torino University. The work of S.F.
has been supported in part by European Community's Human Potential
Program under contract HPRN-CT-2000-00131 Quantum Space-Time, in
association with INFN Frascati National Laboratories and by D.O.E.
grant DE-FG03-91ER40662, Task C. The work of S.V. has been
supported by DFG -- The German Science Foundation, DAAD -- the
German Academic Exchange Service.

\end{document}